\documentclass{pos}

\usepackage{epsfig}
\usepackage{graphicx}

\title{Old nuclear symmetries and large $N_c$ as long distance symmetries
in the two nucleon system}

\ShortTitle{Old nuclear symmetries and large $N_c$ in the two nucleon system}

\author{\speaker{E. Ruiz Arriola}
\thanks{Supported by the
  Spanish DGI and FEDER funds with grant FIS2008-01143/FIS, Junta de
  Andaluc{\'\i}a grant FQM225-05, and EU Integrated Infrastructure
  Initiative Hadron Physics Project contract RII3-CT-2004-506078.}\
\\ Departamento de F\'{\i}sica At\'omica, Molecular y
Nuclear, Universidad de Granada, E-18071 Granada, Spain\ E-mail:
\email{earriola@ugr.es}}

\author{A. Calle Cord\'on\\
        \\ Departamento de F\'{\i}sica At\'omica, Molecular y
Nuclear, Universidad de Granada, E-18071 Granada, Spain\\
        E-mail: \email{alvarocalle@ugr.es}} 

\abstract{Wigner and Serber symmetries for the two-nucleon system
  provide unique examples of long distance symmetries in Nuclear
  Physics, i.e. symmetries of the meson exchange forces broken only at
  arbitrarily small distances. We analyze the large $N_c$ picture as a
  key ingredient to understand these, so far accidental, symmetries
  from a more fundamental viewpoint. A set of sum rules for NN
  phase-shifts, NN potentials and coarse grained $V_{\rm low k}$ NN
  potentials can be derived showing Wigner SU(4) and Serber symmetries
  not to be fully compatible everywhere. The symmetry breaking pattern
  found from the partial wave analysis data, high quality potentials
  in coordinate space at long distances and their $V_{\rm low k}$
  relatives is analyzed on the light of large $N_c$ contracted
  $SU(4)_C$ symmetry. Our results suggest using large $N_c$ potentials
  as long distance ones for the two-nucleon system where the meson
  exchange potential picture is justified and known to be consistent
  with large $N_c$ counting rules. We also show that potentials based
  on chiral expansions do not embody the Wigner and Serber symmetries
  nor do they scale properly with $N_c$. We implement the One Boson
  Exchange potential realization saturated with their leading $N_c$
  contributions due to $\pi,\sigma,\rho$ and $\omega$ mesons. The
  short distance $1/r^3$ singularities stemming from the tensor force
  can be handled by renormalization of the Schr\"odinger equation. A
  good description of deuteron properties and deuteron electromagnetic
  form factors in the impulse approximation for realistic values of
  the meson-nucleon couplings is achieved.}

\FullConference{International Workshop on Effective Field Theories:
  from the pion to the upsilon\\ 2-6 February 2009\\ Valencia, Spain}

\begin{document}

\section{Introduction}

The standard point of view in Particle Physics has often been that
increasing the energy implies a higher degree of symmetry. In QCD, for
instance, scale invariance roughly sets in for momenta much higher
than the quark masses. In Nuclear Physics the situation may be exactly
the opposite; some symmetries such as those introduced by
Wigner~\cite{Wigner:1936dx} and Serber~\footnote{There is no
  reference. According to R. Serber~\cite{Serber:1994ns} the name
  "Serber force" was coined by E. Wigner around 1947.} are unveiled at
low energies where the wavelength becomes larger than a certain
scale. For obvious reasons we call them {\it Long Distance Symmetries}
(LDS)~\cite{CalleCordon:2008cz,Cordon:2009ps}. In the meson exchange
picture this implies the presence of arbitrarily large symmetry
breaking counterterms. We analyze these, so far accidental, LDS in the
two-nucleon system below pion production threshold corresponding to CM
momenta $p \le 400 {\rm MeV}$.

\section{Wigner symmetry}

\begin{figure}
\begin{center}
\epsfig{figure=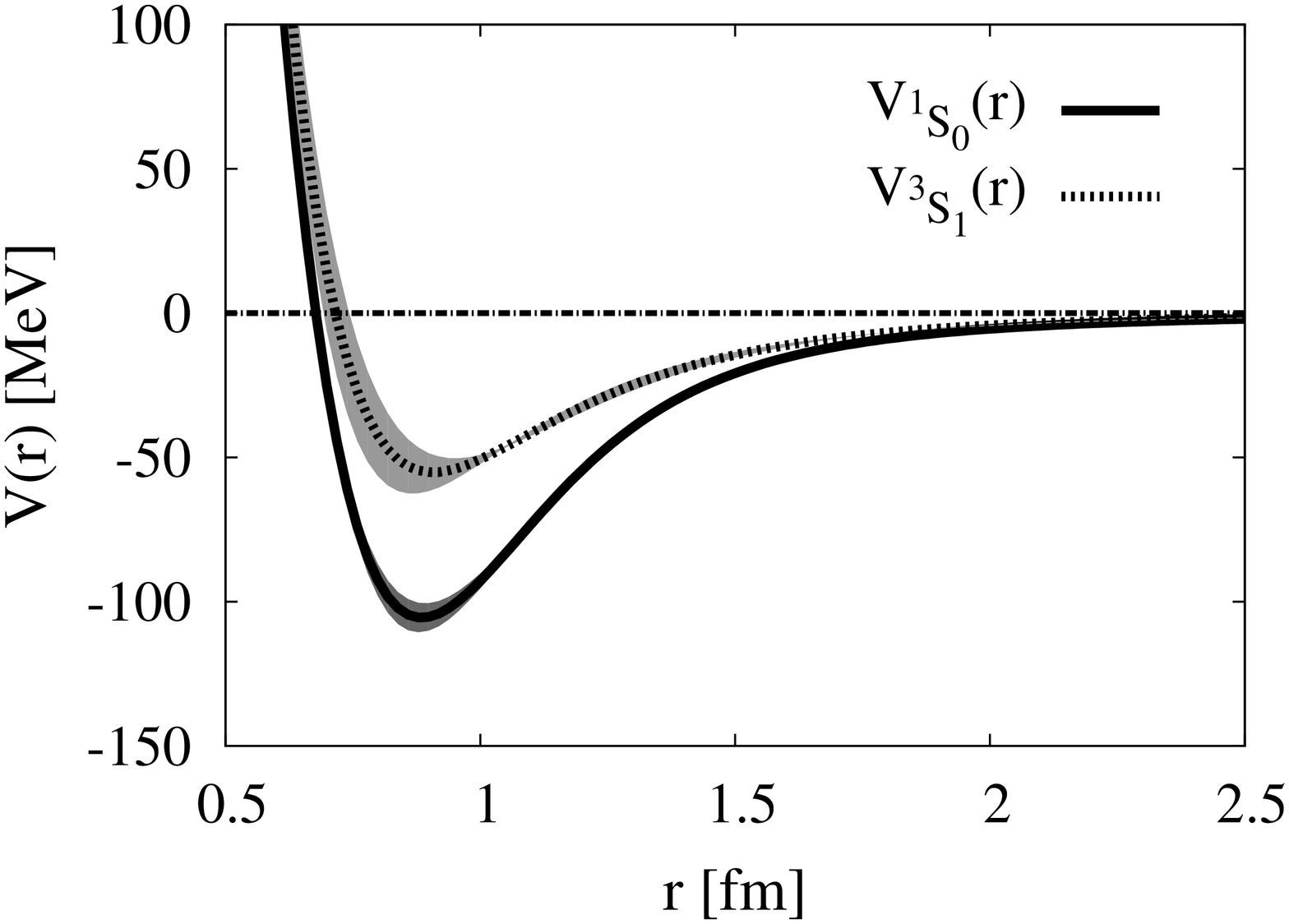,height=5cm,width=4.5cm,angle=270} 
\epsfig{figure=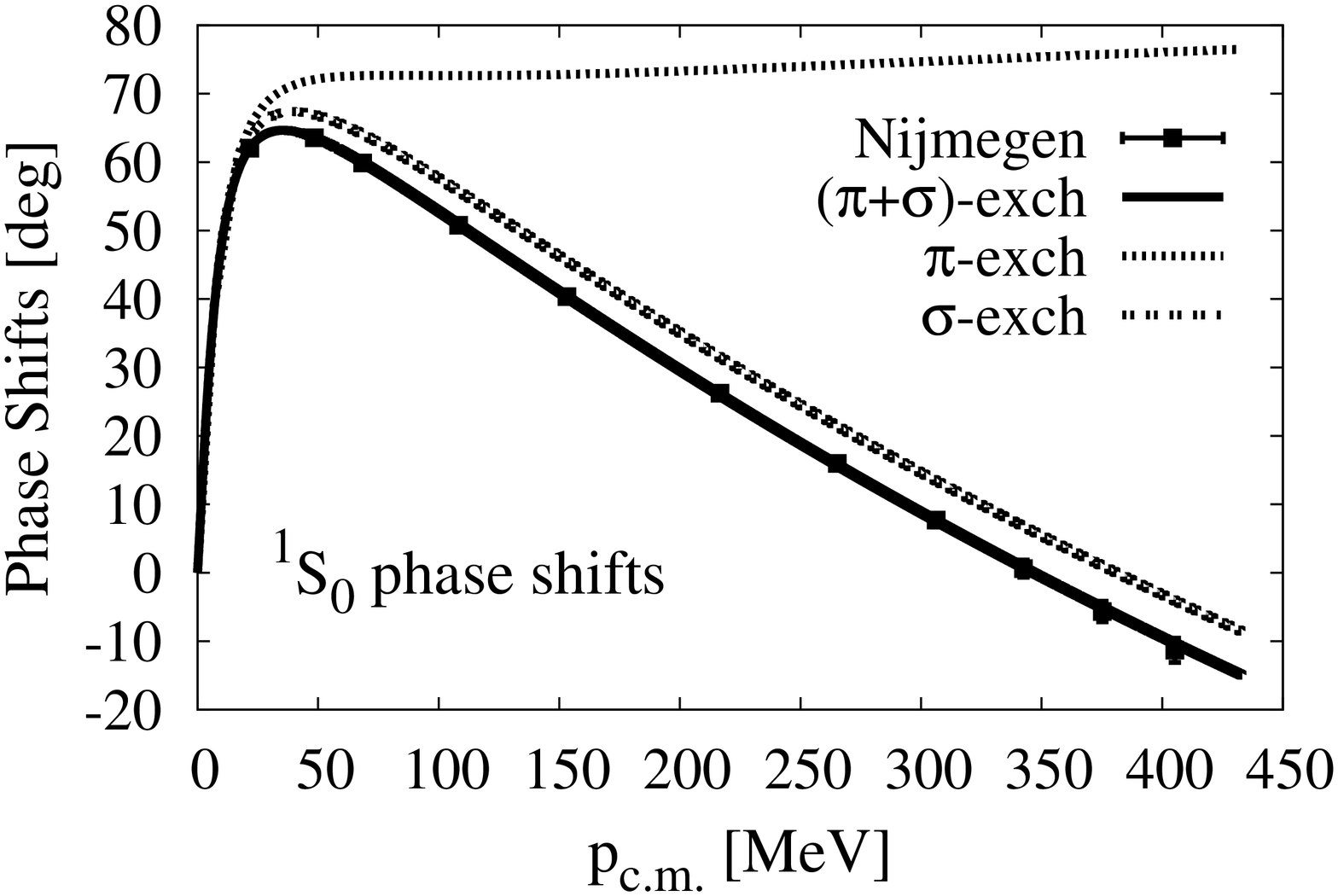,height=4.5cm,width=4.8cm,angle=270}  
\epsfig{figure=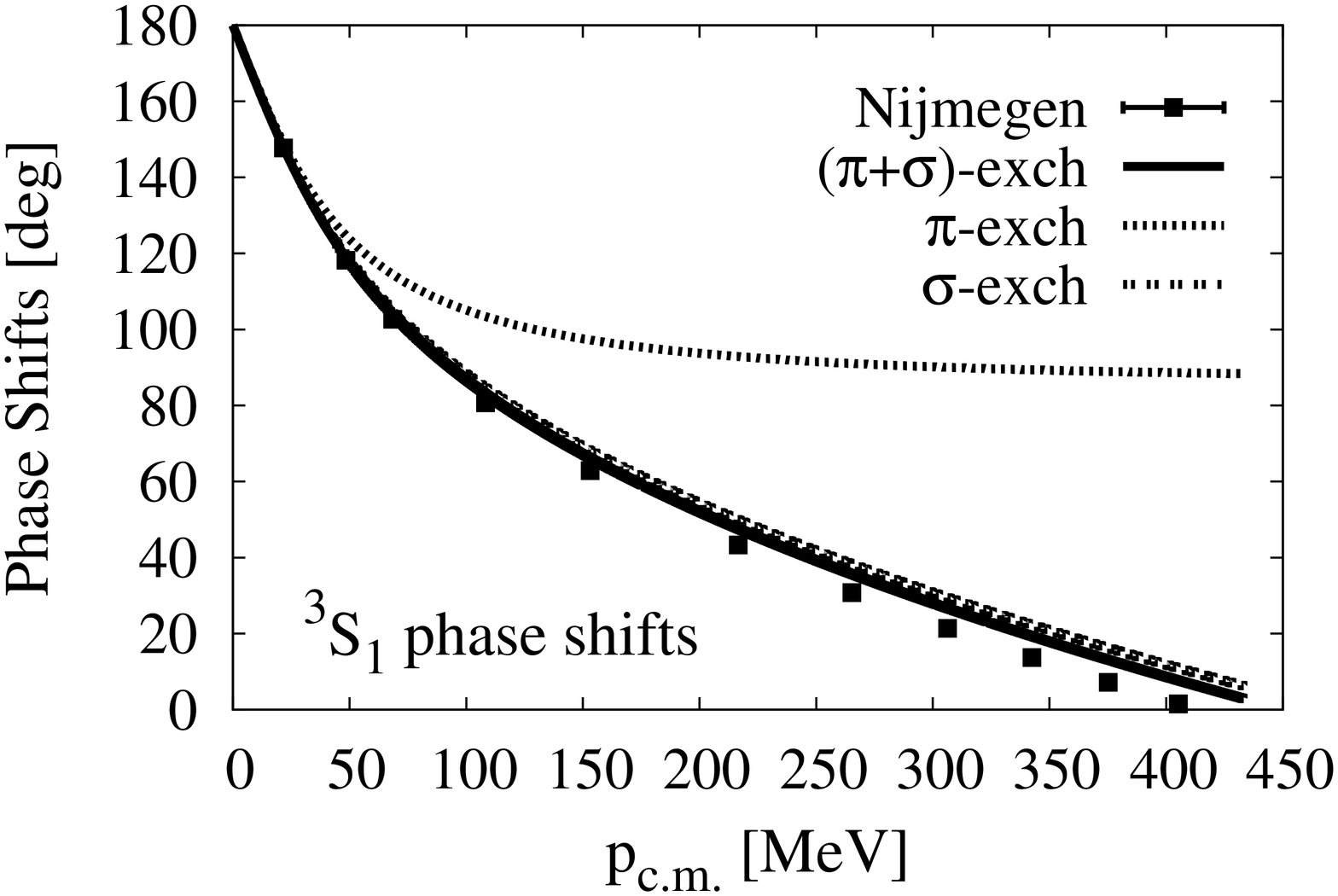,height=4.5cm,width=4.8cm,angle=270}
\end{center}
\caption{Wigner symmetry. Left:NN potentials in the $^1S_0$ and
  $^3S_1$ channel.  Middle:$^1S_0$ NN phase shift. Right:$^3S_1$ NN
  phase shift. Potentials are similar for $r > 2 {\rm fm}$ but
  phase shifts are different {\it everywhere}.}
\label{fig:wigner}
\end{figure}

The Wigner SU(4) spin-flavour symmetry corresponds to the algebra of
isospin $T^a$, spin $S^i$ and Gamow-Teller $G^{ia}$ generators in
terms of the one particle spin $\sigma_A^i$ and isospin $\tau_A^a$
Pauli matrices,
\begin{eqnarray} 
T^a =\frac12 \sum_A\tau_A^a \, , \quad S^i &=& \frac12 \sum_A
\sigma_A^i \, , \quad  G^{ia} = \frac12 \sum_A \sigma_A^i \tau_A^a \, . 
\end{eqnarray} 
The two-body Casimir operator is $ C_{SU(4)} = T^a T_a + S^i S_i +
G^{ia} G_{ia}$. The one-nucleon irreducible representations is a
quartet made of a spin and isospin doublet
$$ {\bf 4}= (p\uparrow, p\downarrow, n\uparrow, n\downarrow)
=(S=1/2,T=1/2) \, .  
$$ Two nucleon states with relative angular momentum $L$ and total
spin $S$ and isospin $T$ fulfilling $(-1)^{S+L+T}=-1$ due to Fermi
statistics correspond to an antisymmetric sextet and a symmetric
decuplet which, in terms of $(S,T)$ representations of the $SU_S(2)
\otimes SU_T(2)$ subgroup, are
\begin{eqnarray} 
{\bf 6}_A &=& (1,0) \oplus(1,0) \quad L=0,2, \dots \quad \to (^1S_0,
^3S_1), (^1D_2, ^3D_{1,2,3}) , (^1G_2, ^3G_{1,2,3}) , \dots \\ {\bf
  10}_S &=& (0,0) \oplus(1,1) \quad L=1,3, \dots \quad \to
\phantom{(^1S_0, ^3S_1),} \, (^1P_1, ^3P_{0,1,2}), (^1F_1, ^3F_{0,1,2}),
\dots
\end{eqnarray} 
In particular, one obtains $V_{^3S_1} (r) = V_{^1S_0}(r)$ which seems
verified for $r > 2 {\rm fm}$ (see Fig.~\ref{fig:wigner}, left) for
high quality potentials~\cite{Stoks:1994wp}, i.e. having $\chi^2 /{\rm
  DOF } < 1 $ for 6000 data !. However, one might think that since a
symmetry of the potential implies a symmetry of the S-matrix one
should also have $\delta_{^1S_0} (p) = \delta_{^3S_1} (p)$ at low
energies, in total contradiction to the data in
Fig.~\ref{fig:wigner}. (see Sect.~\ref{sec:longdist}).  

\section{Serber symmetry}

\begin{figure}
\begin{center}
\epsfig{figure=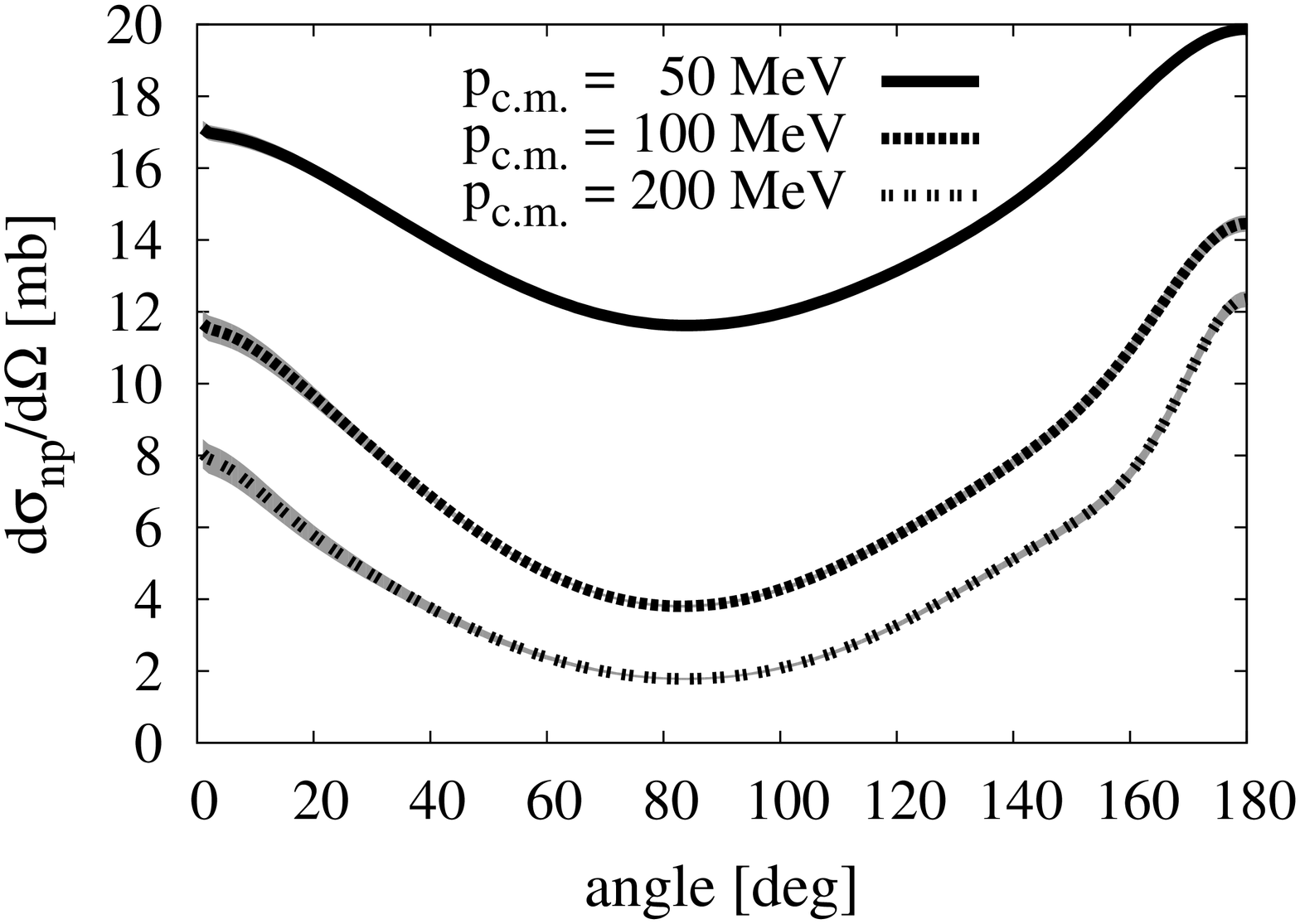,height=4.5cm,width=4.5cm,angle=270}
\epsfig{figure=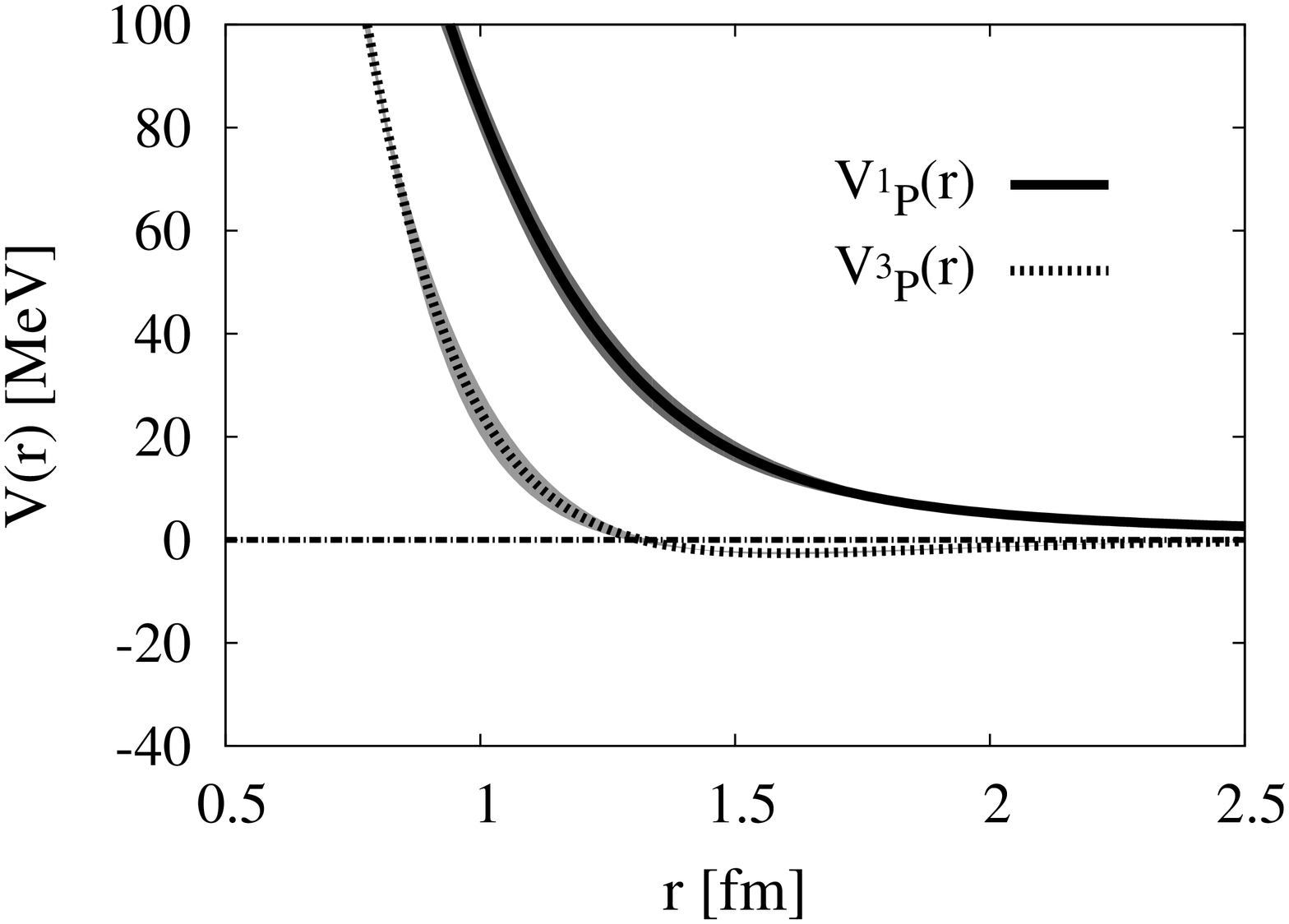,height=5cm,width=4.5cm,angle=270} 
\epsfig{figure=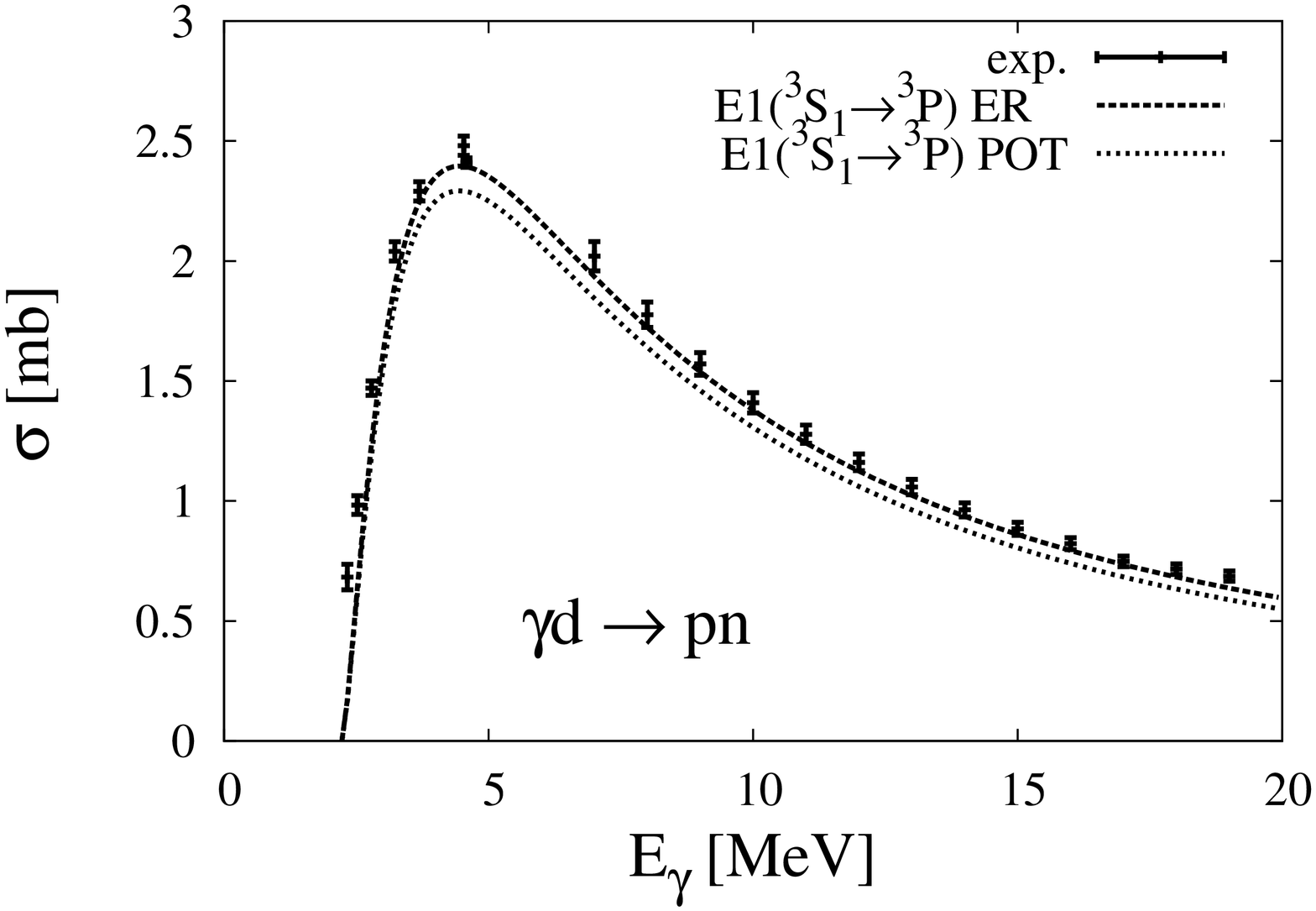,height=5cm,width=4.7cm,angle=270}
\end{center}
\caption{Serber symmetry. Left: pn differential cross section taken
  from an average of the Nijmegen database (nn.online.org). Middle: P-wave spin
  singlet and spin triplet potentials. Right: Deuteron
  photodisintegration as a function of the photon energy based on
  vanishing $^3P$-wave interactions.}
\label{fig:serber}
\end{figure}

A vivid demonstration of Serber symmetry is demonstrated in
Fig.~\ref{fig:serber} (left) where the pn differential cross section
at low CM momenta, $p \le 250 {\rm MeV}$, fulfills to a good
approximation
\begin{eqnarray} 
\frac{d \sigma_{pn}}{d \Omega}= | f_{pn} (\pi-\theta) |^2 = | f_{pn}
(\theta)|^2 \, , 
\end{eqnarray} 
suggesting no interaction in odd L-waves as $P_L( \theta) = (-)^L P_L(
\pi-\theta)$, a fact verified by NN potentials in the spin-triplet
states for $r > 1.2 {\rm fm}$, see Fig.~\ref{fig:serber} (middle) for
the P-wave case. This assumption can also be tested by looking at
Deuteron photodisintegration, $\gamma d \to p n$, dominated above
threshold by the $E_1$ transition $^3S_1 \to ^3P$. Neglecting tensor
force and meson exchange currents (MEC) the cross section for a
normalized deuteron state $u_d(r)$ with binding energy $B_d$
reads~\cite{Arenhovel:1990yg}
\begin{eqnarray} 
\sigma_{E1} ( \gamma d \to pn ) = \frac{\alpha\pi}{3 p} (p^2+ 2
\mu_{pn} B_d) \Big| \int_0^\infty dr u_d (r) \, r \, u_{^3P}(r)
\Big|^2
\end{eqnarray} 
with $E_\gamma = B_d + p^2 /(2 \mu_{pn})$. For a free spherical P-wave
$ u_{^3P}(r) = p r j_1 (pr) $, the agreement is good using $u_d(r)$
from effective range (ER) theory \cite{Arenhovel:1990yg} or from a
potential \cite{CalleCordon:2008cz} (POT), see Fig.~\ref{fig:serber}
(right).

A further hint for Serber symmetry comes from the late 50's Skyrme
proposal~\cite{Skyrme:1959zz} to introduce a pseudopotential 
representing the NN effective interaction in nuclei in the form
\begin{eqnarray} 
V_{\rm effective} ({\bf p}',{\bf p}) &=&   
t_0 (1 + x_0 P_\sigma ) + t_1 (1 + x_1 P_\sigma )
  ({\bf p}'^2 + {\bf p}^2)  + t_2 (1 + x_2 P_\sigma )
        {\bf p}' \cdot {\bf p} + \dots
\label{eq:skyrme2}
\end{eqnarray} 
with $P_\sigma = (1+ \sigma_1 \cdot \sigma_2)/2$  the spin exchange
operator. $P_\sigma=-1$ for spin singlet $S=0$ and $P_\sigma=1$
for spin triplet $S=1$ states. Serber symmetry corresponds to take
$x_2=-1$ in the P-wave term, ${\bf p}' \cdot {\bf p}$. Mean field
theory calculations fitting single nucleon states yield $x_2
=-0.99$~\cite{Zalewski:2008is}.

\section{Renormalization and Long Distance Symmetry}
\label{sec:longdist}

In the meson exchange picture~\cite{Machleidt:1989tm} the NN
interaction can be decomposed as the sum
\begin{eqnarray} 
V(x) = V_{\rm short} (x) + V_{\rm long} (x) 
\end{eqnarray}
where the short range and scheme dependent piece is given by
distributional contact terms 
\begin{eqnarray} 
V_{\rm short} (r) &=& C_0 \delta (\vec x) + C_2 \{ \nabla^2 , \delta
(\vec x) \} + \dots  \, , 
\end{eqnarray}
whereas the long distance piece $V_{\rm long} (x)$ is scheme
independent and usually produces power divergences $\sim 1/r^n $ at
short distances. We introduce a short distance
cut-off,$r_c$, which will be removed in the end~\footnote{The
  constants $C_0$, $C_2$ etc. are scale dependent. The equivalence
  with momentum space renormalization is shown in
  Ref.~\cite{Entem:2007jg} where the limit $r_c \to 0$ implies the
  irrelevance of $C_2$ in the presence of a singular chiral
  potential.}.  LDS means that even if $V_{^1S_0}(r)=V_{^3S_1}(r)$ for
any $r > r_c $ one has $C_{0,^1S_0} \neq C_{0,^3S_1}$.  We analyze the
implications by looking at finite energy $S-$wave scattering states
\begin{eqnarray}
u_p (r) = u_{p,c} (r) + p \cot \delta_0 (p) \, u_{p,s} (r) \to \cos (p
r) + \cot \delta_0 (p) \sin (p r) \, ,  
\end{eqnarray} 
where $p = \sqrt{2\mu_{pn}E} $ is CM momentum.  For $p \to 0$ then
$\delta_0 (p) \to - \alpha_0 p $ and zero energy states are
\begin{eqnarray}
u_0 (r) = u_{0,c} (r)  - u_{0,s} (r) / \alpha_0 \to 1-r/\alpha_0    \, ,
\end{eqnarray} 
Here $u_{p,c} (r) $, $ u_{p,s} (r)$, $u_{0,c} (r) $ and $ u_{0,s}
(r)$ depend on $V(r)$ {\it only}. Orthogonality in $r_c \le r <
\infty$ requires
\begin{eqnarray}
0= \int_{r_c}^\infty dr \left[u_{0,c} (r) - \frac{1}{\alpha_0} \,
u_{0,s} (r) \right] \Big[ u_{p,c} (r) + p \cot
\delta_0 (p) \, u_{p,s} (r) \Big] \, . 
\label{eq:ort} 
\end{eqnarray} 
Note that the potential $V(r)$ and the scattering length $\alpha_0$
are {\it independent variables}. Thus we assume Wigner symmetry for
the potential $V_{^1S_0} (r) = V_{^3S_1}(r) $ but experimentally
different scattering lengths $\alpha_{^1S_0}= -23.74 {\rm fm} $ and $
\alpha_{^3S_1}= 5.42 {\rm fm} $, yielding from Eq.~(\ref{eq:ort}) the
structure for $r_c \to 0 $,
\begin{eqnarray}
p \cot \delta_{^1S_0} (p) = \frac{ \alpha_{^1S_0} {\cal A} ( p) +
{\cal B} (p)}{ \alpha_{^1S_0} {\cal C} ( p) + {\cal D} (p)} \, , \qquad p
\cot \delta_{^3S_1} (p) = \frac{ \alpha_{^3S_1} {\cal A} ( p) + {\cal
B} (p)}{ \alpha_{^3S_1} {\cal C} ( p) + {\cal D} (p)} \, , 
\end{eqnarray} 
showing that a symmetry of the potential for any $r > r_c$, $r_c \to 0
$, {\it is not} necessarily a symmetry of the S-matrix.  The result
for $\pi+\sigma$ exchange, while not exact, works rather well (see
Fig.~\ref{fig:wigner}).

\begin{figure}[]
\begin{center}
\epsfig{figure=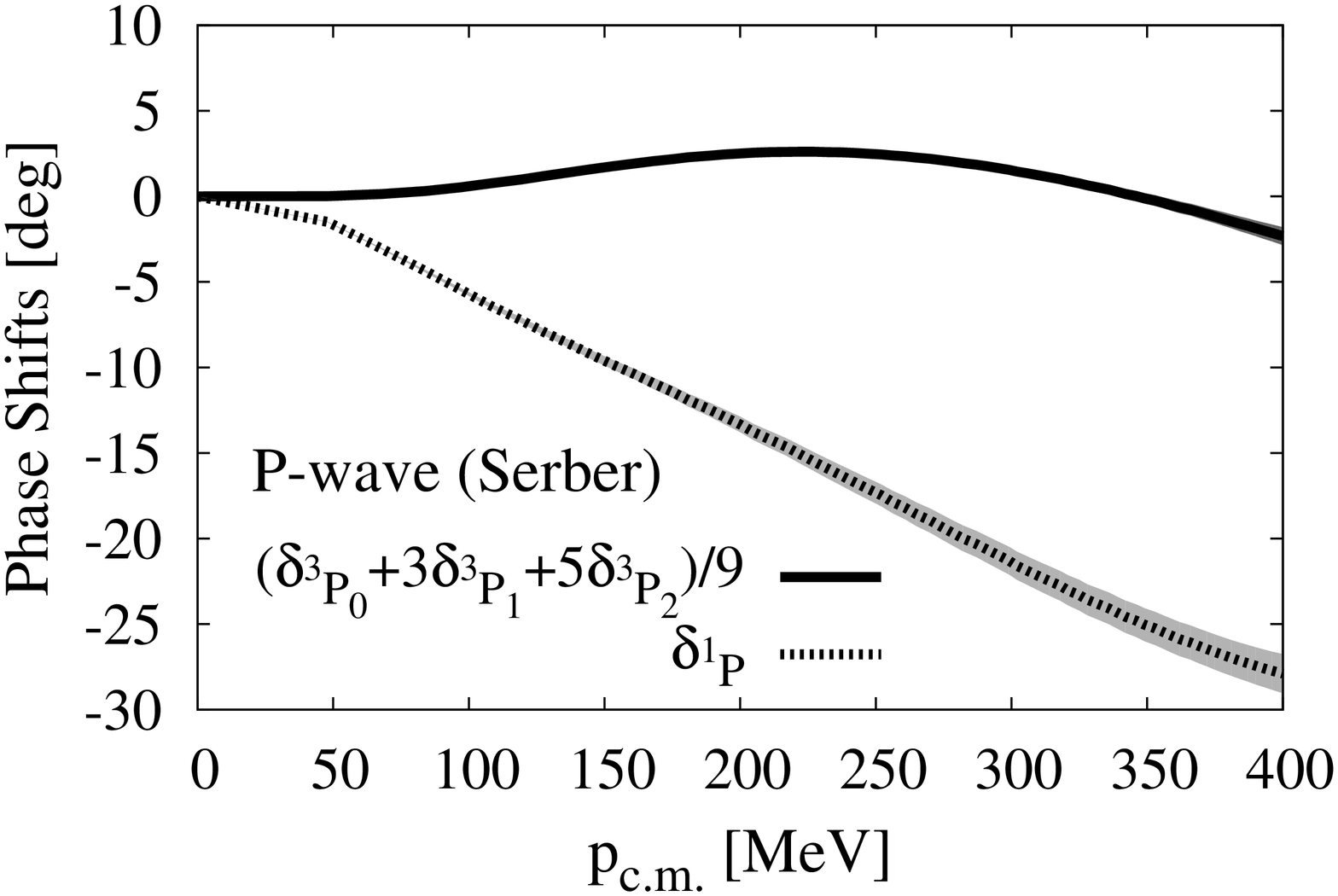,height=3.6cm,width=3.8cm,angle=270}
\epsfig{figure=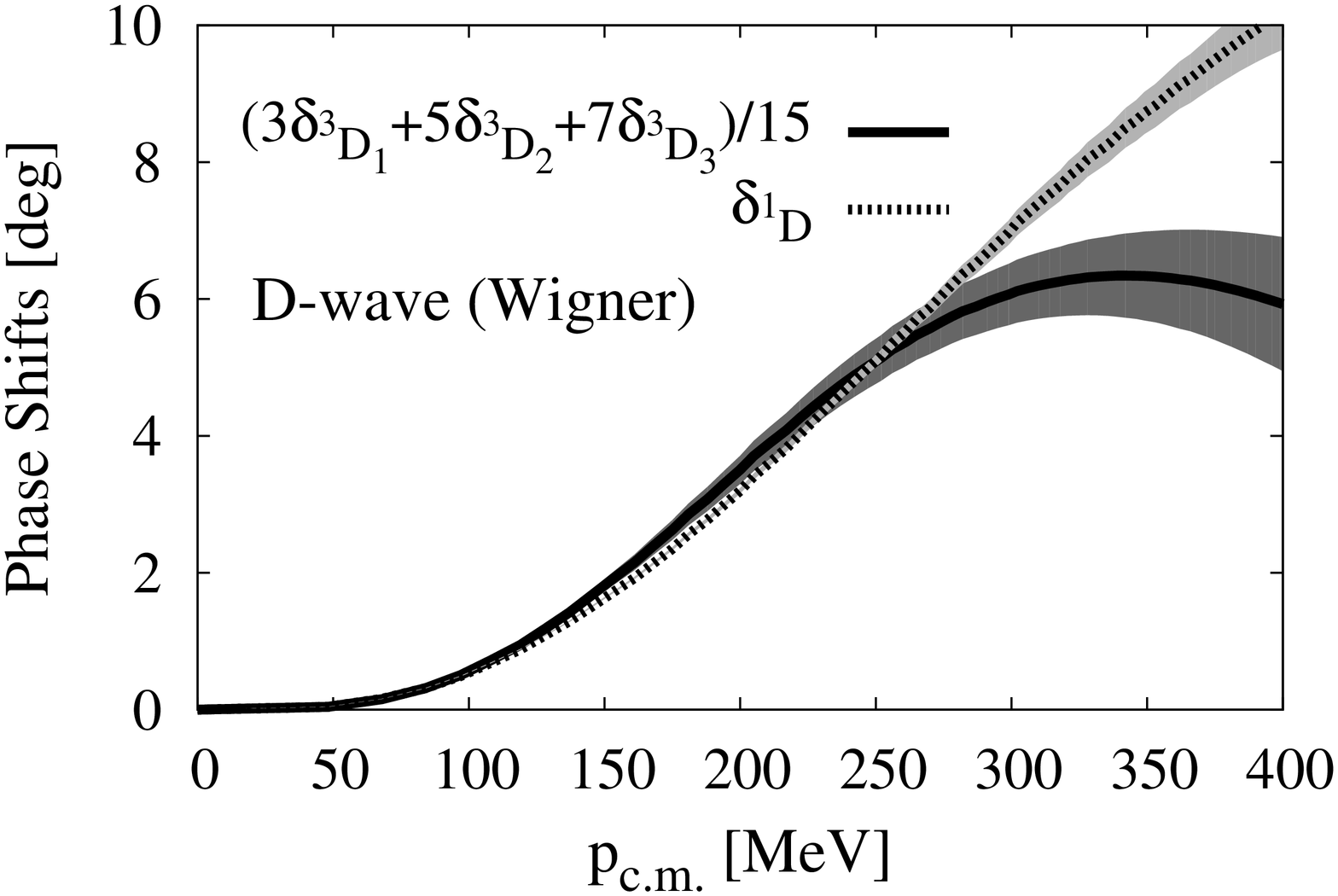,height=3.6cm,width=3.8cm,angle=270}
\epsfig{figure=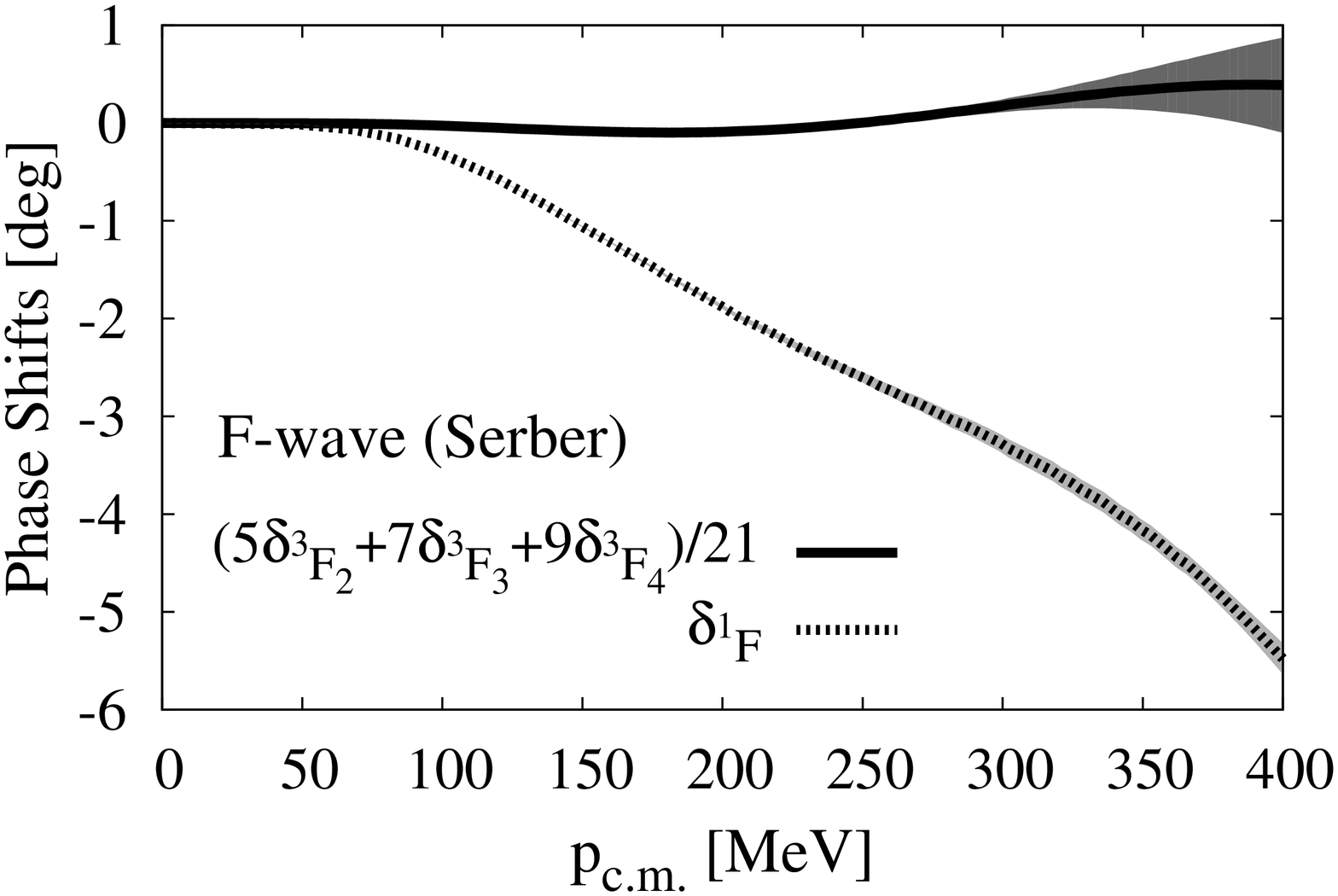,height=3.6cm,width=3.8cm,angle=270}
\epsfig{figure=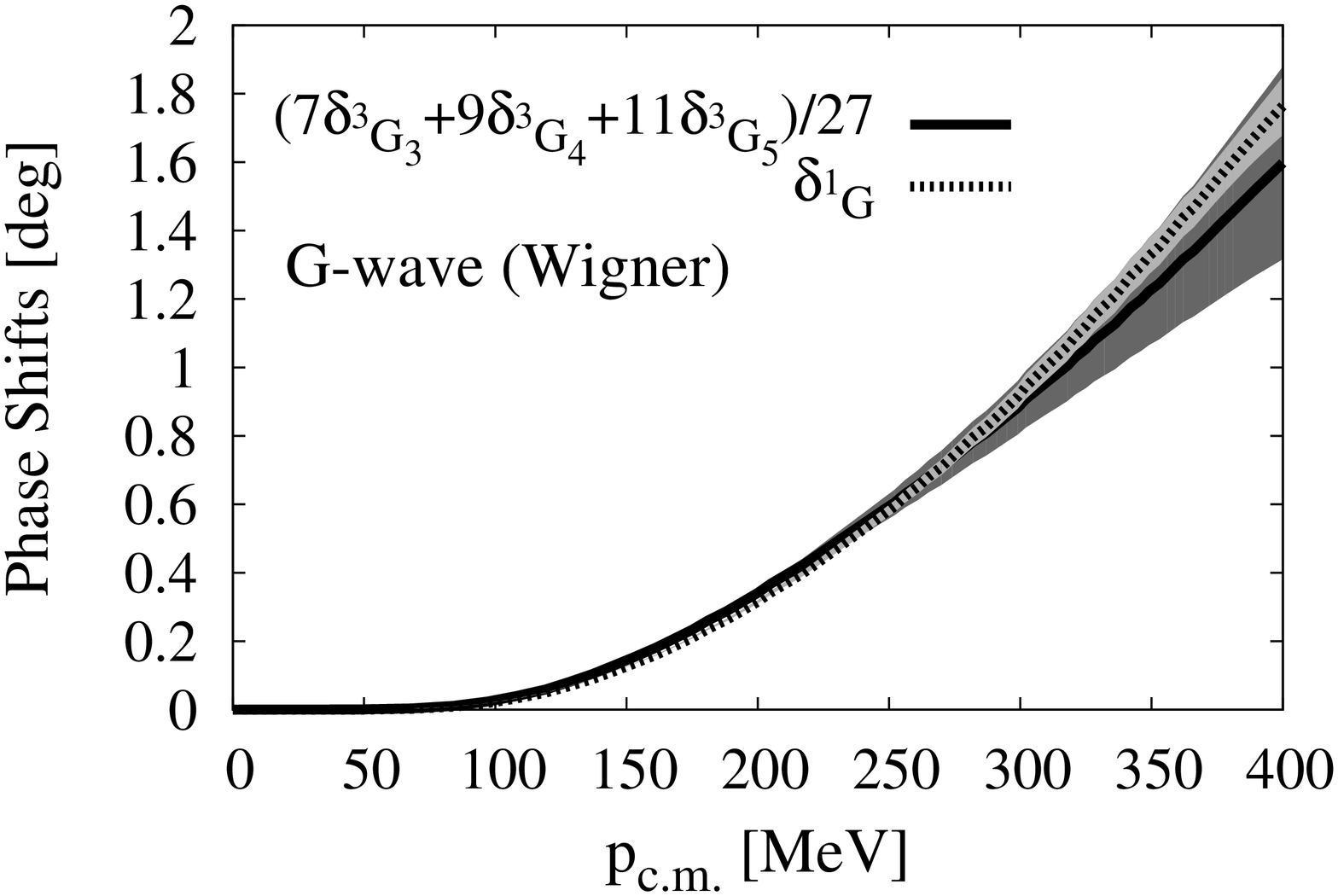,height=3.6cm,width=3.8cm,angle=270}
\end{center}
\caption{Phase shifts sum rules based on spin-orbit and tensor force
  breaking to first order. Wigner symmetry is verified only by even-L
  states while Serber symmetry is verified only by spin triplet odd-L
  waves.}
\label{fig:sumrules}
\end{figure}

\section{Sum rules}

Based on the LDS idea 
we have recently derived the sum rules
for phase shifts~\cite{CalleCordon:2008cz,Cordon:2009ps}
\begin{eqnarray}
\underbrace{\delta_{^3L} (p)= \delta_{^1L} (p) \qquad {\rm all} \,\, 
  L}_{\rm Wigner} \qquad , \qquad 
\underbrace{\delta_{^3L} (p) , \delta_{^1L} (p) = 0 \qquad {\rm odd}
  \, \, L}_{\rm Serber} \quad , 
\end{eqnarray} 
where we have defined the multiplet center $\delta_{L}^{ST} =
1/(3(2L+1)) \sum_{J=L-1}^{L+1} (2J+1) \delta_{LJ}^{ST} $.  From data
Fig.~\ref{fig:sumrules} shows that one has Wigner for {\it even} L and
Serber for triplet {\it odd} L. The LDS character accommodates the
symmetry for increasing $p$ and $L$; what matters is the impact
parameter, $b \sim L/p$.

The previous sum rules have a parallel long distance potential analog
, and are also well verified for $r > 1.5 {\rm
  fm}$~\cite{Cordon:2009ps}. This suggests that a coarse graining of
the interaction using e.g. the $V_{\rm low k}$
potentials~\cite{Bogner:2003wn} works and justifies {\it per se} the
symmetry obtained phenomenologically by fitting single particle
states~\cite{Zalewski:2008is} for the Skyrme effective force,
Eq.~(\ref{eq:skyrme2}),~\cite{Cordon:2009ps}.  We find that $
V_{^3L, {\rm low k}}(p,p) \ll V_{^1L, {\rm low k}}(p,p)$ for
$(-1)^L=-1$ and $V_{^3L, {\rm low k}}(p,p) \sim V_{^1L, {\rm low
    k}}(p,p)$ for $(-1)^L=1$.

\section{Large $N_c$ nucleon-nucleon potentials}

\begin{figure}[ttt]
\begin{center}
\epsfig{figure=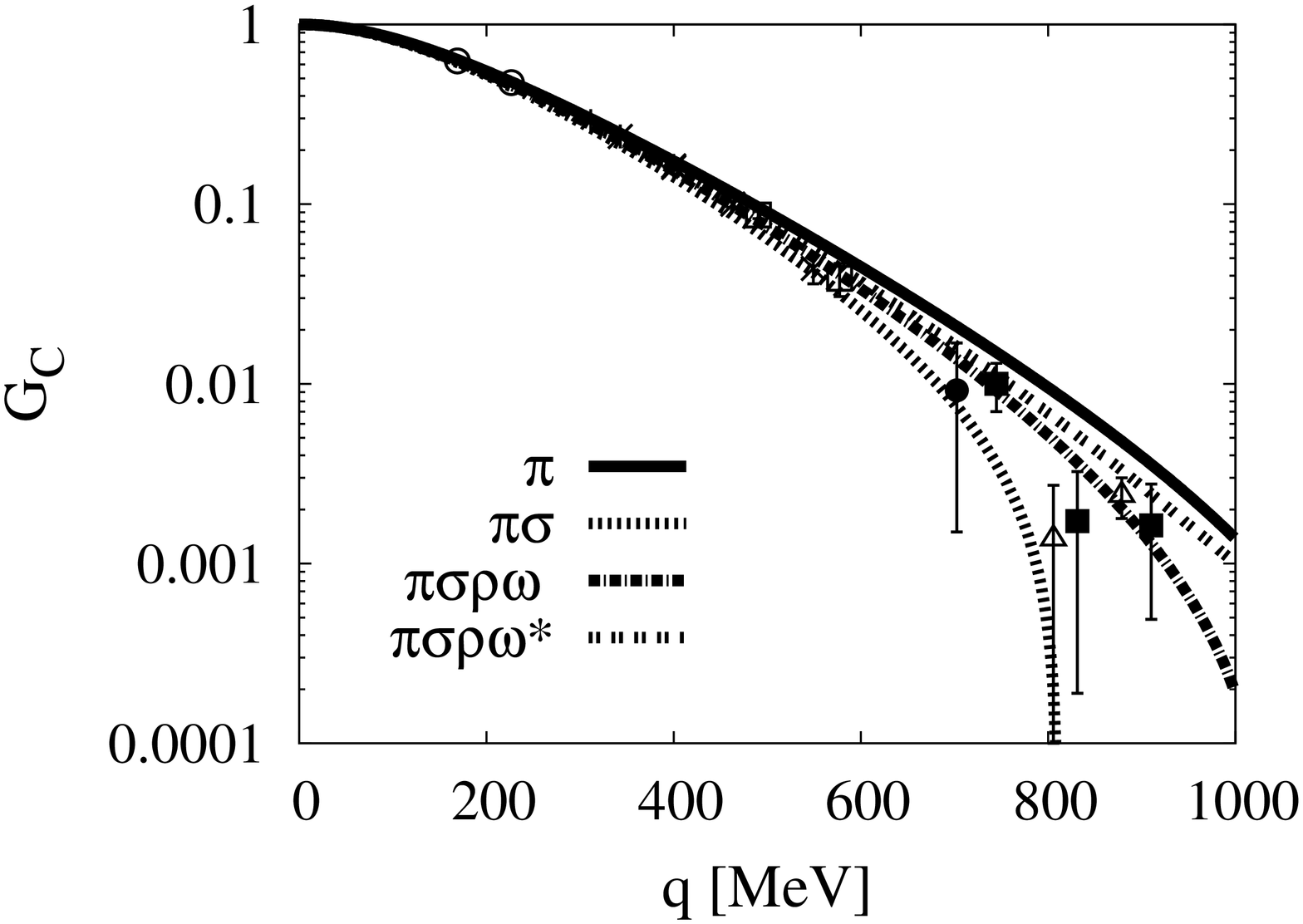,height=4.7cm,width=4.1cm,angle=270}
\epsfig{figure=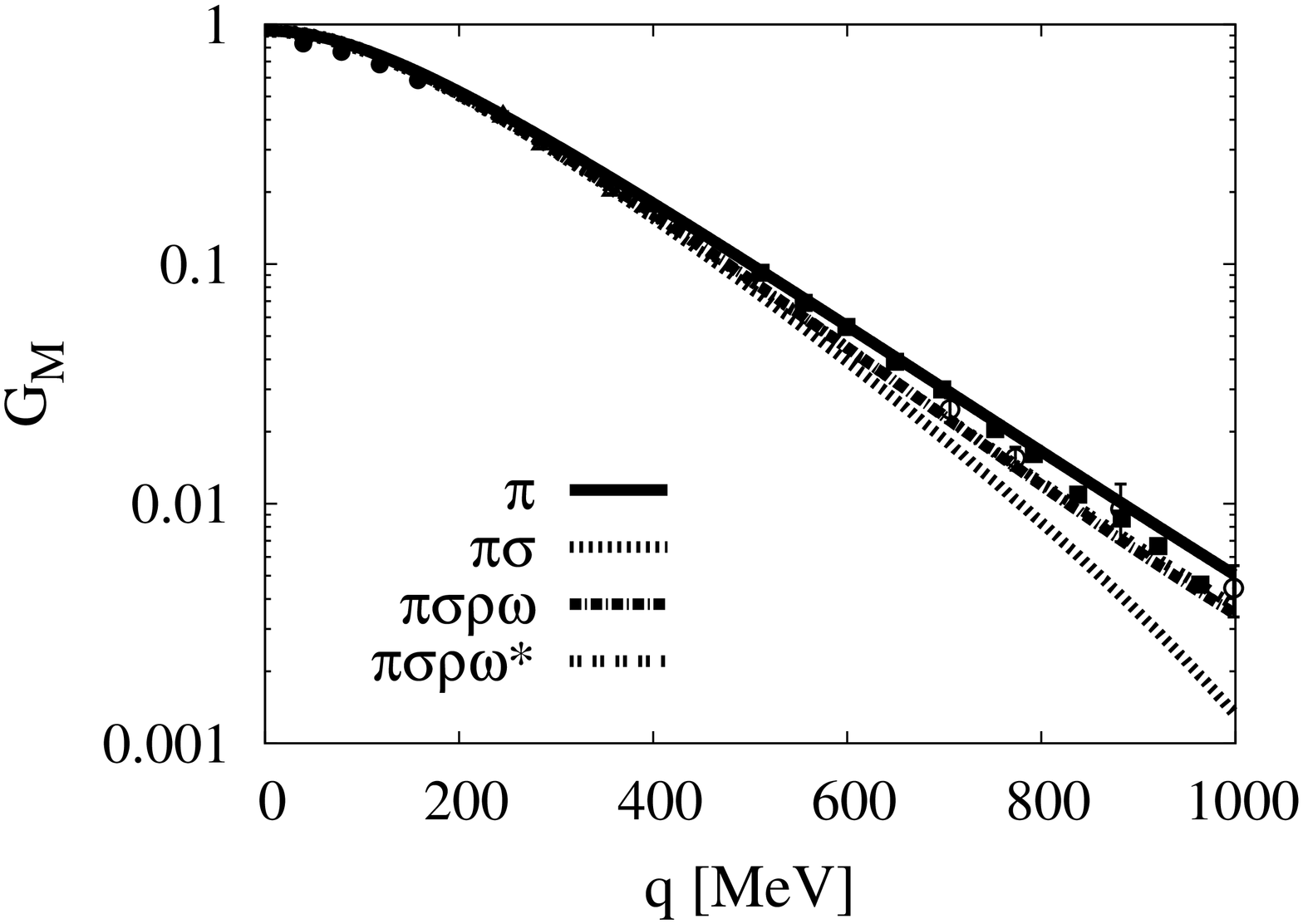,height=4.7cm,width=4.1cm,angle=270}
\epsfig{figure=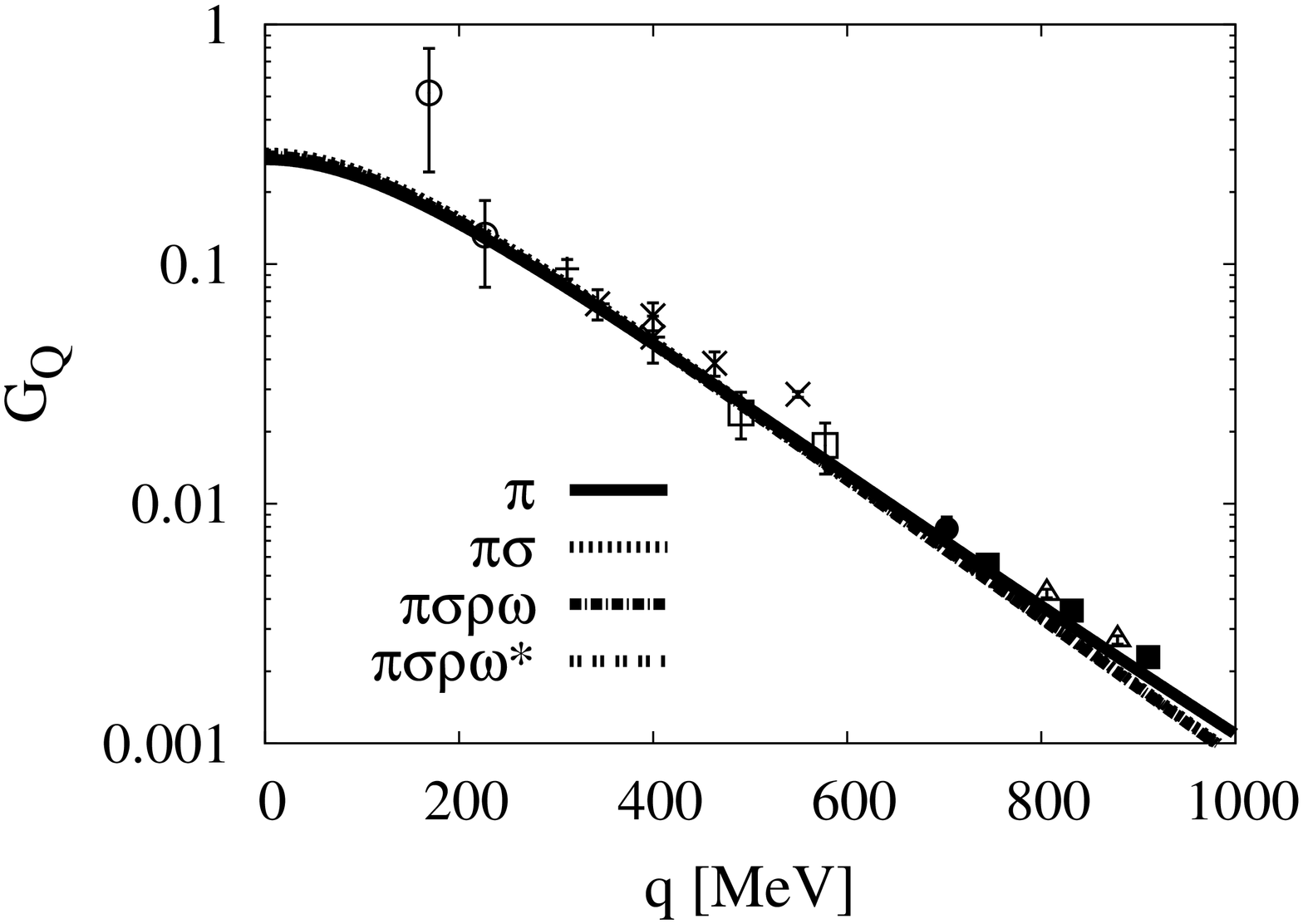,height=4.7cm,width=4.1cm,angle=270}
\end{center}
\caption{Deuteron charge (left), magnetic (middle) and quadrupole
  (right) form factors. See also Table 1.}
\label{fig:formfactors}
\end{figure}

As it is well known, in the large $N_c$ limit with $\alpha_s N_c$ fixed,
nucleons are heavy, $M_N \sim N_c$~\cite{Witten:1979kh}, and the NN
potential $\sim N_c$ becomes meaningful. The amazing aspect is that
the symmetry pattern of the sum rules for the old nuclear Wigner and
Serber symmetries largely complies to the large $N_c$ and QCD based
contracted $SU(4)_C$ symmetry~\cite{Kaplan:1995yg,Kaplan:1996rk} where
the tensorial spin-flavour structure is
\begin{eqnarray} 
V (r) = V_C (r) + \tau_1 \cdot \tau_2 [ \sigma_1 \cdot \sigma_2  W_S (r)
+ S_{12}  W_T(r) ]  \sim N_c 
\end{eqnarray} 
Other operators are ${\cal O} (N_c^{-1}) $ and hence suppressed by a
relative $1/N_c^2$ factor. One has the sum rules
\begin{eqnarray}
V_{^1L} (r)= V_{^3L} (r)&=& V_C (r) - 3 W_S (r) +{\cal O} (N_c^{-1})
\, , \quad (-1)^L=+1 \\ V_{^1L} (r) &=& V_C (r) + 9 W_S (r) +{\cal O}
(N_c^{-1}) \, , \quad (-1)^L=-1 \\ V_{^3L} (r) &=& V_C (r) + \, \,
\,W_S (r) + {\cal O} (N_c^{-1}) \, , \quad (-1)^L=-1
\end{eqnarray} 
Thus, large $N_c$ {\it implies} Wigner symmetry {\it only} in even-L
channels, exactly as observed in Fig.~\ref{fig:sumrules}. Serber
symmetry is possible but less evident (see \cite{Cordon:2009ps}).
This suggests to use large $N_c$ itself and its contracted
spin-flavour group $SU(4)_C$ as a long distance symmetry.  Actually,
the energy independent potential may be obtained in a multi-meson
exchange picture consistently with large $N_c$ counting
rules~\cite{Banerjee:2001js}~\footnote{The LDS character implies
  relaxing the contact interaction piece {\it not} to be of the same
  form as the long distance potentials, i.e. $V_{\rm short}(\vec x)
  \neq (C_C + \tau_1 \cdot \tau_2 [ \sigma_1 \cdot \sigma_2 C_S +
    S_{12} C_T ] ) \delta (\vec x)$ avoiding the extra symmetry,
  $\tau_a \sigma_i \to -\tau_a
  \sigma_i$~\cite{Cohen:2003eb}.}. Retaining one boson exchange (OBE)
with $\pi$,$\sigma$,$\rho$ and $\omega$ mesons one has
\begin{eqnarray}
V_C (r) &=& - \frac{g_{\sigma NN}^2}{4 \pi} \frac{e^{-m_\sigma r}}{r}
+ \frac{g_{\omega NN}^2}{4 \pi} \frac{e^{-m_\omega r}}{r}  \, , 
\label{eq:vc} \\ 
W_S(r)
&=& \frac{g_{\pi NN}^2}{48\pi} \frac{m_\pi^2}{\Lambda_N^2}
\frac{e^{-m_\pi r}}{r} + \frac{f_{\rho NN}^2}{24
\pi}\frac{m_\rho^2}{\Lambda_N^2} \frac{e^{-m_\rho r}}{r}  \, , 
\\ 
W_T(r) &=&
\frac{g_{\pi NN}^2}{48 \pi}\frac{m_\pi^2}{\Lambda_N^2} \frac{e^{-m_\pi r}}{r} \left[ 1 + \frac{3}{m_\pi r} + \frac{3 }{(m_\pi r)^2}\right] 
- \frac{f_{\rho NN}^2}{48 \pi}\frac{m_\rho^2}{\Lambda_N^2} \frac{e^{-m_\rho
r}}{r} \left[ 1 + \frac{3}{m_\rho r} + \frac{3 }{(m_\rho r)^2}\right] \, ,  
\end{eqnarray} 
where $\Lambda_N = 3 M_p /N_c$ and $g_{\sigma NN}, g_{\pi NN}, f_{\rho
  NN}, g_{\omega NN} \sim \sqrt{N_c}$ and $m_\pi , m_\sigma, m_\rho,
m_\omega \sim N_c^0$. To leading and subleading order in $N_c$ one may
neglect spin orbit, meson widths and relativity.  The tensor force
$W_T$ is singular at short distances $\sim 1/r^3$ and requires
renormalization (see \cite{PavonValderrama:2005gu} for the $\pi$
case). Deuteron properties are shown in Table~\ref{tab:table_triplet}
for parameters always reproducing the $^1S_0$ phase shift,
Fig.~\ref{fig:wigner} (middle).  Space-like electromagnetic form
factors in the impulse approximation~\cite{Gilman:2001yh} for $G_E^p
(-{\bf q}^2)=1/(1+{\bf q}^2/m_\rho^2)^2$ and without MEC are plotted
in Fig.~\ref{fig:formfactors} (see \cite{Valderrama:2007ja} for the
$\pi$ case). Overall, the agreement is good for {\it realistic}
couplings~\footnote{The Goldberger-Treiman relation gives $g_{\pi NN}=
  g_A M_N /f_\pi = 12.8 $ for pions and $g_{\sigma NN}= M_N /f_\pi =
  10.1 $ for scalars for $f_\pi=92.3 {\rm MeV}$ and $g_A=1.26$.
  Sakurai's universality and KSFR yield $ g_{\rho NN} = g_{\rho \pi
    \pi}/2 = m_\rho / f_\pi /\sqrt{8} = 2.9$. From $SU(3)$ we have $
  g_{\omega NN} =3 g_{\rho NN} - g_{\phi NN} = 8.7$ using OZI rule,
  $g_{\phi NN} =0$. $\rho-$meson dominance yields $ f_{\rho NN}
  = \kappa_\rho g_{\rho NN} $ with $ \kappa_\rho= \mu_p - \mu_n -1
  =3.7$ with $\mu_p=2.79$ and $\mu_n=-1.91$. Adding 
  $\rho',\rho''$ states yields $\kappa_\rho=6.1$ and thus $f_{\rho
    NN}=18$.}. The inclusion of shorter range mesons induces moderate
changes, due to the expected short distance insensitivity embodied by
renormalization, {\it despite} the short distance singularity and {\it
  without} introducing strong meson-nucleon-nucleon vertex
functions. In practice convergence is achieved for $r_c \sim 0.3 {\rm
  fm}$. Our calculation includes only the OBE part of the leading
$N_c$ potential but multiple meson exchanges could also be
added~\cite{Banerjee:2001js}.

\begin{table}[]
\caption{Deuteron properties for renormalized large $N_c$ OBE
  potentials. We use $ \gamma= \sqrt{ 2 \mu_{np} B_d} $ with
  $B_d=2.224575(9)$ and take $g_{\pi NN} =13.1083 $, $m_\pi=138.03
  {\rm MeV} $, $m_\rho=770 {\rm MeV} $, $m_\omega=782 {\rm MeV} $.  A
  fit to the $^1S_0$ phase shift gives $m_\sigma=501 {\rm MeV}$ and
  $g_{\sigma NN}=9.1$~\cite{CalleCordon:2008eu}. $\pi\sigma\rho\omega$
  uses $f_{\rho NN}=15.5$ and $g_{\omega NN}=9.857$ while
  $\pi\sigma\rho\omega^*$ uses $f_{\rho NN}=17.0$ and $g_{\omega
    NN}=10.147$. Experimental or recommended values from
  Ref.~\cite{deSwart:1995ui}.}
\begin{tabular}{|c|c|c|c|c|c|c|c|}
\hline & $\gamma ({\rm fm}^{-1})$ & $\eta$ & $A_S ( {\rm fm}^{-1/2}) $
& $r_m ({\rm fm})$ & $Q_d ( {\rm fm}^2) $ & $P_D $ & $\langle r^{-1}
\rangle $  \\ \hline
$\pi$(\cite{PavonValderrama:2005gu}) & Input & 0.02633 & 0.8681 & 1.9351 & 0.2762
& 7.88\% & 0.476  \\ 
\hline 
$\pi\sigma$ & Input & 0.02599 & 0.9054 & 2.0098 & 0.2910
& 6.23\% & 0.432  \\ 
\hline 
$\pi\sigma\rho\omega$ & Input & 0.02597 & 0.8902 & 1.9773 & 0.2819
& 7.22\% & 0.491  \\ 
$\pi\sigma\rho\omega$$^*$ 
& Input & 0.02625 & 0.8846 & 1.9659 & 0.2821
& 9.09\% & 0.497  \\ 
\hline 
NijmII(\cite{Stoks:1994wp}) & Input & 0.02521 & 0.8845(8) & 1.9675 & 0.2707 & 
5.635\% &  0.4502  \\
Reid93(\cite{Stoks:1994wp}) & Input & 0.02514 & 0.8845(8) & 1.9686 & 0.2703 & 
5.699\% & 0.4515  \\ \hline 
%
Exp. (\cite{deSwart:1995ui}) &  0.231605 &  0.0256(4)  & 0.8846(9) & 1.9754(9)  &
0.2859(3) & 5.67(4) &  \\ \hline 
\end{tabular}
\label{tab:table_triplet}
\end{table}


For large $N_c$, the central potential is leading, Eq.~(\ref{eq:vc}).
Energy independent potentials using power counting within Chiral
Perturbation Theory (ChPT)~\cite{Kaiser:1997mw} yield a central force
$V_C^{\rm ChPT}$ only to ${\cal O} ( 1/f_\pi^4 M_N) $ i.e. N$^2$LO and
ChPT potentials do not scale properly with $N_c$ since $g_A \sim N_c$,
$f_\pi \sim \sqrt{N_c}$ and there are terms scaling as $ V_{\rm
  2\pi}^{\rm ChPT} \sim g_A^4 / f_\pi^4 \sim N_c^2 $ and not as $\sim
N_c$, even after inclusion of $\Delta$~\cite{Kaiser:1998wa}. Moreover,
Wigner and Serber symmetries are violated at long distances since
\begin{eqnarray} 
V_{\rm 2\pi}^{\rm ChPT} (r) = ( 1 + 2 \tau_1 \cdot \tau_2 ) \frac{e^{-2m_\pi
    r}}{r} \frac{3 g_A^4 m_\pi^5}{1024 f_\pi^4 M_N \pi^2}+ \dots 
\end{eqnarray} 
These features might perhaps explain why renormalizing ChPT potentials
in different schemes a mismatch of $10^0$ at $p=400 {\rm MeV}$ for the
$^1S_0$ phase shift is persistently
obtained~\cite{PavonValderrama:2005wv,Higa:2007gz,
  Entem:2007jg,Valderrama:2008kj}.

\section{Conclusions}

Wigner and Serber symmetries in the NN system are realized as long
distance ones and are largely compatible with the large $N_c$
picture. When large $N_c$ NN-potentials are saturated by
$\pi$,$\sigma$,$\rho$ and $\omega$ exchange and subsequently
renormalized, we obtain satisfactory results for the deuteron and
central partial waves.  This suggests that large $N_c$ potentials
might eventually provide a workable scheme, less directly related to
ChPT but closer in spirit to the common wisdom of Nuclear Physics.

\end{document}